\documentstyle{ichep}
\def\tbar{{\bar{t}}}
\def\tt{t\bar{t}}
\def\to{\rightarrow}

\def\qq{q\bar{q}}
\def\pp{{\rm p\bar{p}}}
\def\bbar{{\bar{b}}}
\def\qbar{{\bar{q}}}
\def\bb{{b\bar{b}}}
\def\ww{W^+W^-}
\def\GeV{{\rm GeV}}

\def\TeV{{\rm TeV}}
\def\gt{\Gamma_t}
\def\bWbW{b W^+ \bar b W^-}
\def\degree{^{\circ}}
\def\cF{{\cal F}}
\def\prod{\cF_{\mbox{\tiny PROD}}}
\def\dec{\cF_{\mbox{\tiny DEC}}}
\def\int{\cF_{\mbox{\tiny INT}}}

\begin{document}

\title{
\vskip -3.5\baselineskip
\begin{flushright}
{\normalsize\rm
UR-1379 \\
 DTP/94/72  \\
ER-40685-829 \\
August 1994 \\}
\end{flushright}
\vskip 2\baselineskip
Soft Jets and Top Mass Measurement at the Tevatron}

\author{Lynne H Orr$^{\dag\ddag\#}$\ and W J Stirling$^{\S\|}$}

\affil{\dag\ Department of Physics and Astronomy, University of Rochester,\\
Rochester, NY 14627, USA\\
\S\ Departments of Physics and Mathematical Sciences, University of Durham,\\
Durham DH1 3LE, UK}

\abstract{
Extra soft jets in top events in $p \bar p$ collisions may arise not only
from gluons radiated off initial state partons or final state $b$ quarks, but
may also be radiated from the $t$ quarks themselves.  We discuss predictions
for distributions of soft gluons in $t \bar t$ production at the Tevatron
and the implications for attempts to measure the top mass by reconstructing
the invariant mass of its decay products.}

\twocolumn[\maketitle]

\fnm{7}{Presented by L H Orr at the XXVII International Conference on High
Energy Physics, Glasgow, July 20-27, 1994; proceedings to be published by
IOP Publishing Ltd.}
\fnm{2}{Work supported in part by the U.S. DOE, grant DE-FG02-91ER40685.}
\fnm{8}{E-mail: orr@urhep.pas.rochester.edu.}
\fnm{4}{E-mail: wjs@hep.durham.ac.uk.}

\section{Introduction}
In $\tt$ production at the Tevatron, the final state particles may be
accompanied by additional soft jets due to gluon radiation.  These soft jets
must be accounted for somehow in attempts to measure the top mass $m_t$ by
momentum reconstruction.  In particular, one would like to know whether
soft jets should be combined with the top's daughter $W$'s and $b$'s in
such reconstructions.  It is obvious that if the gluon has been radiated
off the final $b$ or $\bbar$, the gluon should be included, but if
it was radiated off an intitial state quark, then it should not.  Our
intuition tells us that final-state radiation, as in the former case,
 corresponds to jets near the $b$ or
$\bbar$ direction, and that initial-state radiation, as in
the latter case, corresponds to jets near the beam axis.

This intuitive picture is incomplete, however, because we must also consider
radiation off the top quarks themselves.  Do such gluons belong to the inital
state or the final state?  That this question cannot be answered indicates
that the the initial/final state picture of gluon radiation is too na{\"\i}ve
in the case of the top quark.  Top production and decay must be considered
simultaneously in a treatment of gluon radiation.

In this talk we report results of a study \cite{OS} of soft gluon radiation in
top production and decay in which all diagrams are correctly taken into
account.
Our aims are (i) to determine where the gluons come from and where they go, in
a way that is relevant to $m_t$ measurement, and (ii) to compare the correct
results which those of simple, intuitive models that are in the spirit of
what might be easily implemented in Monte Carlo Simulations.

\section{Soft gluons: formalism and features}

\begin{figure*}
\vspace{14cm}
\vspace{7cm}
\hspace{-2.2cm}
\hspace{-.2in}
%\special{psfile=ichep1.ps}
\vspace{-14.cm}
\caption{
Gluon pseudorapidity distributions in $\tt$ production via
$\qq\to \tt \to \bb \ww $, in $\pp$ collisions at
$\protect\sqrt{s} = 1.8\ \TeV$.
(a) Net distribution and contributions from
production and decay.
(b) Distributions arising from ISR, ISR/FSR, and BB models described in
the text.}
\vspace{-.23cm}
\end{figure*}

We work in the soft approximation ({\it i.e.}, we assume that the gluons
are less energetic than other particle in the event); for a discussion of
the soft gluon formalism in top physics see \cite{KOS}.  We consider
the process $\qq\to\tt\to\bWbW$ with emission of a soft gluon.
The matrix element and phase space factorize so that we can write the
gluon distribution as
\begin{equation}
{1\over d\sigma_0}\ {d\sigma\over{d E_g d\cos\theta_g d\phi_g}}\ = \
{\alpha_s\over 4 \pi^2} \ E_g \ (\prod + \dec +\int),
\end{equation}
where $d\sigma_0$ is the differential cross section for the lowest-order
process (with no gluon).
$\prod$ corresponds to gluons radiated in
association with $\tt$ production, {\it i.e.,} radiated before the $t$ or
$\tbar$ quark goes on shell.
  Similarly, $\dec$ corresponds to gluons
radiated in the decay of the $t$ or $\tbar$.  $\int$ represents the
interferences between the two and depends on the top width $\gt$.
Expressions for $\prod$, $\dec$, and $\int$ can be found in \cite{KOhS}.

The important point is that this production--decay--interference decomposition
provides a gauge--invariant substitute for the initial/final state picture
discussed above.  It
determines for us whether the gluon's momentum should be combined with those
of the $t$ decay products in reconstructing the top quark's four-momentum.
Gluons associated with top production do not contribute to the on-shell top
quark's momentum and should not be included.  Gluons associated with the
decay {\it do} contribute to the top momentum.  For the interference term there
is no such clear interpretation, but in the case of interest here it is
negligible anyway.

Detailed discussions of the properties of this distribution
(Eq.\ 1) can be found in
\cite{OS} and \cite{KOhS}.  Here we merely wish to point out some
physical features of $\prod$ and $\dec$ which have consequences for the
full distributions we see below, and which distinguish the correct distribution
from those in simpler models.  Being associated with $\tt$ production only,
$\prod$ knows nothing about the decay of the top quark, and depends only
on the momenta of the initial $q$ and $\qbar$ and the $t$ and $\tbar$,
as well as that of the gluon itself.  Similarly, $\dec$ knows nothing about
the initial state and depends only on the momenta of the $t$, $\tbar$, $b$,
$\bbar$, and gluon.  Both $\prod$ and $\dec$ can be written as sums of
``color antennae'' which can be interpreted in terms of a pair of quarks
connected by a color string.  These antennae exhibit color coherence, or
the string effect:  more radiation appears between such paired quarks than
outside of them.

\begin{figure*}
\vspace{14cm}
\vspace{7cm}
\hspace{-2.2cm}
\hspace{-.2in}
%\special{psfile=ichep2.ps}
\vspace{-14.cm}
\caption{Distribution in the cosine of the angle between the gluon and the
$b$-quark,  (a) with cuts described in the text and
 (b) with the additional cut $|\eta_g|\leq 1.5$.}
\vspace{-.23cm}
\end{figure*}

\section{Gluon distributions at the Tevatron}
Let us examine soft gluon distributions for $\tt$ production in $\pp$
collisions
at 1.8 TeV center-of-mass energy at the Tevatron.  The results shown are
from \cite{OS}, where a more complete discussion can be found.
We take $m_t=174\ \GeV$
and work at the parton level, considering only the $\qq$ initial state (which
dominates) and using minimal kinematic cuts, which are:
%($|\eta_{b}|,|\eta_{\bbar}|  \leq  1.5;\  |\eta_g| \leq 3.5;\
%10\ \GeV/c \leq p_T^g  \leq  25\ \GeV/c;\
%E_g \leq 100\ \GeV$; and $\Delta R_{bg},\Delta R_{\bbar g} \geq 0.5$).
\begin{eqnarray}
|\eta_{b}|,|\eta_{\bbar}| \> & \leq & \> 1.5 \; ,\nonumber \\
|\eta_g| \> & \leq & \> 3.5 \; ,\nonumber \\
10\ \GeV/c \leq \> & p_T^g & \> \leq  25\ \GeV/c \; ,\nonumber \\
E_g \> & \leq & \> 100\ \GeV \; ,\nonumber \\
\Delta R_{bg},\Delta R_{\bbar g} \> & \geq & \> 0.5 \; .
\label{cuts}
\end{eqnarray}

\subsection{Angular distributions and top momentum reconstruction}
We focus on angular distributions since we are interested in where soft jets
will appear in detectors.  Figure 1(a) shows the gluon pseudorapidity
distribution.  The total (solid line) is shown along with its decomposition
according to Eq.\ 1 into
production (dotted line) and decay (dashed line) contributions.  The
production piece is peaked in the forward direction and centrally suppressed.
This reflects the color antennae connecting the initial-state quarks with the
top quarks.  The decay contribution is peaked in the central region, which
is where the radiating top and bottom quarks tend to be produced.  The net
distribution is only slightly peaked in the center.
Note that,
while gluons at larger rapidities are almost exclusively associated with
production (and hence should be ignored in top momentum reconstruction),
central gluons are nearly as likely to have come from production as from decay.

In Figure 2 we test the second part of our guess (see introduction) by
 examining  to what extent proximity of gluons to the $b$ quark
correlates with having come from the decay contribution.  Fig. 2(a)
shows the distribution in cosine of the angle between the $b$ quark and the
gluon with the same cuts and decomposition as in Fig. 1(a).
The contribution from production is flat, as expected since it contains
no explicit  dependence on the $b$ quark's momentum.  The decay
contribution does increase as the gluon approaches the $b$, leading to
an excess above the production contribution close to the $b$.
The excess is only a slight one, though, and
the result is very sensitive to the cut on $\Delta R$.  Furthermore, no
hadronization effects have been taken into account.  We can improve the
situation by recalling that forward gluons tend to come from production.
If we tighten the gluon pseudorapidity cut to $|\eta_g|<1.5$, we see more
of an excess in decay gluons near the $b$, as shown in Fig.\ 2(b).  Sensitivity
to the $\Delta R$ cut and fragmentation effects remain a problem, however.

\begin{figure*}
\vspace{14cm}
\vspace{7cm}
\hspace{-2.2cm}
\hspace{-.2in}
%\special{psfile=ichep3.ps}
\vspace{-14.cm}
\caption{Forward--backward asymmetry in gluon pseudorapidity distributions in
$\tt$ production.  The cuts are as in Fig.\ 1 with the additional
requirements $\Delta\phi_{\bb}>135\degree$ and $\Delta\phi_{bg}<90\degree$.
The curves correspond to the
(a) total (solid), production (dots) and decay (dashes) distributions, and
to the  distributions for the
(b)  ISR (dots), ISR/FSR (dashes)
and BB (dash-dots) models.}
\vspace{-.23cm}
\end{figure*}

We now return to the pseudorapidity distribution
to compare the correct  distribution in
Fig.~1(a) to those in Fig.~1(b), obtained
from some simpler models that are intuitively appealing and easily implemented
in Monte Carlo simulations.  The ISR model (dotted line)
includes radiation off the initial $\qq$ state only, as if the $q$ and $\qbar$
formed a color singlet.  We might expect this to correspond to the
contribution associated with production, but we see by comparing to the
dotted line in Fig.~1(a) that the ISR model overestimates radiation in the
central region.  In the ISR/FSR model (dashed line) we add to the ISR model
radiation from the final $\bb$
pair as if they too formed a color singlet.  This model corresponds roughly
to the na\"{\i}ve expectation mentioned in the introductory paragraph.
Figure 1(b) shows that
this model overestimates the total radiation and gets the shape wrong.
In the BB model (dot-dashed line) we use the correct color structure but ignore
radiation off the top quark.  This model approximately
reproduces the correct pseudorapidity distribution.  However, it does not
give the correct azimuthal distribution,\cite{OS} and, more important
for $m_t$ reconstruction, does not permit a production--decay decomposition.

\subsection{Color structure and forward-backward asymmetry}

Finally, we discuss briefly a forward-backward asymmetry in the
radiation pattern (for appropriately chosen final states)
that arises from the color
structure of gluon emission in hadronic $\tt$ production.  While not directly
relevant to measurement of the top mass, the asymmetry is interesting because
it is a result of the fact that the top quarks themselves can radiate before
decaying.  It also reveals major differences between the correct distribution
and the simpler models.

This asymmetry arises from the string effect mentioned above.
For example, in $\qq\to\tt$ the $q$--$t$ antenna produces more radiation in the
region between the $t$ and $q$ than, say, between the $t$ and $\qbar$,
resulting in a forward--backward asymmetry in the gluon
radiation.  To avoid cancellation of the effect by an equal and opposite
asymmetry due to the $\qbar$ and $\tbar$, we try to preferentially select
gluons that are more likely to be in the $t$ than the $\tbar$ hemisphere, with
the additional cuts $\Delta \phi_{\bb} > 135\degree$ and
$\Delta \phi_{bg}<90\degree$.

The resulting distribution is shown in Figure
3(a).  A forward--backward asymmetry is evident, and we see from the
decomposition that it comes entirely from the production piece; the decay
knows nothing about the initial quarks' direction.  In Fig.\ 3(b) we show
the same distribution for the three simpler models.  There is no asymmetry
for the ISR and ISR/FSR models because there is no connection between radiation
in the initial and final states.  In contrast, the BB model shows a more
marked asymmetry than the correct distribution because without radiation from
the intermediate top quarks there is a more direct color connection between the
initial and final states.

\section{Summary}

We have shown that the subject of gluon radiation in $\tt$ production and
decay is a complicated one due to the rich color structure of the process.
For purposes of top mass reconstruction, we saw that there is no simple
prescription for dealing with additional soft jets in $\tt$ events, but
that the production--decay decomposition provides some guidance.
A comparison to simpler,
intuitively appealing models such as one might easily
implement in Monte Carlo simulations showed that they
do not reproduce the correct distributions and/or do not allow for the
production--decay decomposition.
Finally, we discussed a forward--backward asymmetry in soft gluon radiation
that illustrates the color structure, including in particular
radiation off the top quarks themselves.

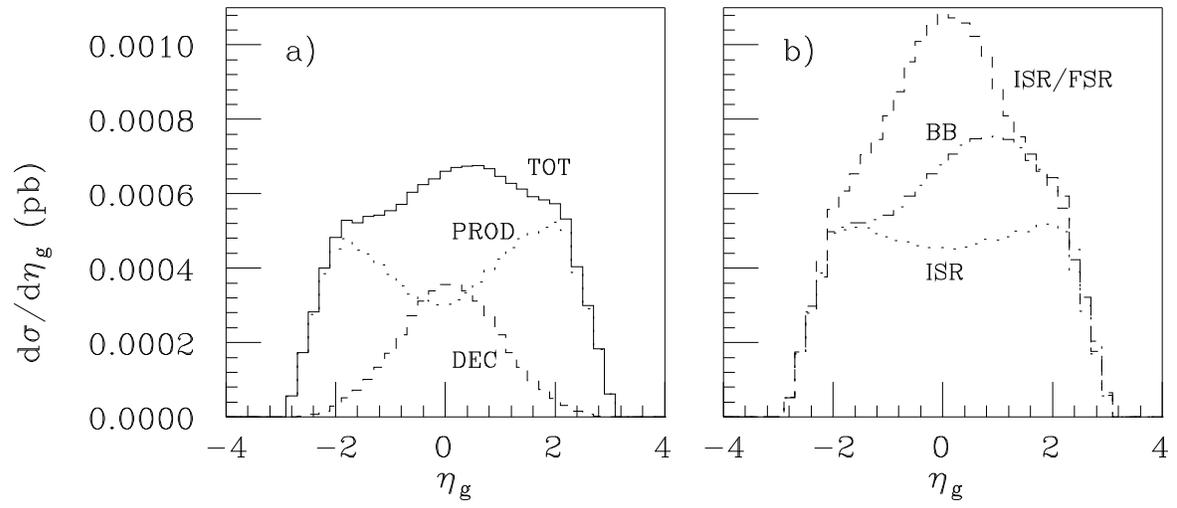
\end{document}